\documentclass[12pt,english,aps,manuscript]{article}
\usepackage[T1]{fontenc}
\usepackage[latin1]{inputenc}
\usepackage{geometry}
\usepackage{amsmath}
\usepackage{amssymb}
\usepackage[numbers]{natbib}

\makeatletter

\usepackage{geometry}

\makeatletter

\usepackage{color}

\makeatletter
\newcommand{\bee}{\begin{equation}}
\newcommand{\ee}{\end{equation}}
\newcommand{\beea}{\begin{eqnarray}}
\newcommand{\eea}{\end{eqnarray}}

\makeatother

\makeatother

\makeatother

\usepackage{babel}
\begin{document}
~~~~~~~~~~~~~~~~~~~~~~~~~~~~~~~~~~~~~~~~~~~~~~~~~~~~~~~~~~~~~~~~~~~~~~~~~~~~~~~~~~~~~~~~~~~~~~~~~~~~~~~~~~~~~~~~~~~~~~~~~~~~~~~~~~~~~~~~~~~~~~~~~~~~COLO-HEP-570

\begin{center}
\textbf{\Large Constraints on LVS Compactifications of IIB String
Theory}{\Large{} }
\par\end{center}{\Large \par}

\begin{center}
\vspace{0.3cm}
 
\par\end{center}

\begin{center}
{\large S. P. de Alwis$^{\dagger}$ } 
\par\end{center}

\begin{center}
Physics Department, University of Colorado, \\
 Boulder, CO 80309 USA 
\par\end{center}

\begin{center}
\vspace{0.3cm}
 
\par\end{center}

\begin{center}
\textbf{Abstract} 
\par\end{center}

We argue that once all theoretical and phenomenological constraints
are imposed on the different versions of the Large Volume Scenario
(LVS) compactifications of type IIB string theory, one particular
version is favored. This is essentially a sequestered one in which
the soft terms are generated by Weyl anomaly and RG running effects.
We also show that arguments questioning sequestering in LVS models
are not relevant in this case.

\begin{center}
\vspace{0.3cm}
 
\par\end{center}

\vfill{}

$^{\dagger}$dealwiss@colorado.edu

\eject

\section{Introduction}

The Large Volume Scenario (LVS) is at this point the only viable framework
for discussing the phenomenology of string theory compactifications.
While Heterotic constructions have advanced further in terms of actually
getting close to the standard model (in terms of the spectrum), getting
a stabilized compactification with broken supersymmetry, with a cosmological
constant that can at least be tuned to a scale far below the scale
of SUSY breaking, has proved to be a daunting task%
\footnote{In string theory there are no fine-tunable parameters. Thus the only
known way of getting solutions with SUSY breaking and a small CC is
by finding appropriate flux configurations. The general argument for
this was given in \citep{Bousso:2000xa}. While this is possible in
type II theories it seems highly unlikely in heterotic constructions
because there is only one type of flux. Papers which claim to derive
a heterotic (or M-theory) phenomenology ignore this point.%
}. On the other hand in the case of type IIB theories, although the
construction of the standard model (without chiral exotics) is still
a work in progress, the problem of getting a SUSY breaking stabilized
vacuum with a small cosmological constant (CC) has already been solved
\citep{Balasubramanian:2005zx}. The two problems are in fact decoupled.
This is a consequence of the fact that the standard model physics
is local, and has only a marginal effect on the global problem of
stabilization, SUSY breaking, and tuning the CC.

However the phenomenological consequences of a flux compactified theory
can never be predictive in any precise quantitative sense. As explained
in \citep{deAlwis:2006nm}, this is a result of the existence of a
landscape of flux compactifications and the tuning of the CC in the
general framework of such compactifications \citep{Bousso:2000xa}.
Nevertheless, assuming that the standard model exists within this
context, it is possible to derive qualitative features of a SUSY breaking
scenario. This is the approach that was taken in our earlier papers
on this subject and will be taken here as well.

The types of LVS models that exist in the literature can be classified
by the volume (${\cal V}$ in string units), the gravitino dependence
of the classical contribution to the soft mass $m_{0}$, and the size
of the string scale.
\begin{itemize}
\item Partially `sequestered' $m_{0}\sim m_{3/2}/\ln m_{3/2}$
\item Sequestered $m_{0}\sim m_{3/2}/\sqrt{\ln m_{3/2}{\cal V}}$
\item TeV scale string theory
\item Moduli mixing case leading to $m_{0}\sim m_{3/2}$.
\end{itemize}
Let us first consider the last two cases.

An LVS scenario which gave a string scale of a TeV or so was recently
presented in \citep{Cicoli:2011yy}. This involves a very large compactification
volume $ $${\cal V}$ of order $10^{30}$ so as to give a string
scale $M_{string}\sim M_{P}/\sqrt{{\cal V}}\sim1TeV$. The usual scenario
where the standard model is on a D3 or D7 brane will not work here,
since this will give soft terms at the milli-electron volt scale with
the gravitino mass $m_{3/2}<M_{P}/10^{30}\sim10^{-3}eV$. One possibility
suggested in this reference is to have the standard model on an anti-D-brane.
However it is far from clear that a viable stable classical configuration
can be obtained in the presence of anti-D-branes since their back
reaction will de-stabilize the system%
\footnote{For discussion of the effect of anti-D branes in the original GKP-KKLT
system see \citep{Brustein:2004xn}.%
}. In any case once an anti-D brane is introduced, there is no way
to describe the physics in terms of four dimensional supergravity.
So it is not at all clear how the standard LVS discussion, which makes
essential use of the supergravity formalism in 4D applies. 

The last possibility, leading to soft masses at the gravitino scale
\citep{Choi:2010gm}, arises if there is moduli mixing as suggested
in \citep{Conlon:2010ji}. However the argument for this relies on
pushing the field theory formalism beyond the string scale. This is
not a valid extrapolation. At energy scales greater than the string
scale, one cannot use a local operator formalism as in field theory
to describe the physics. The $\alpha'$ expansion is clearly invalid
at energies larger than the string scale.

This leaves us with the first two alternatives. The first case (that
of partial sequestering) is the one that was originally suggested
in 2006 \citep{Conlon:2006wz}, but was thought to be non-viable as
a consequence of the observation that it is not possible to stabilize
the standard model brane using fluxes \citep{Blumenhagen:2007sm}.
However this possibility has been revived in a recent paper \citep{Cicoli:2011qg}.
Nevertheless we will show that a combination of theoretical and phenomenological
constraints rules out the case of partial sequestering. Essentially
the problem has to do with the large volume ( ${\cal V}>10^{15}$)
that is necessary in that case to avoid the problem of flavor changing
neutral currents. Such a large volume clearly will not lead to a viable
phenomenology. We are left with the second alternative \citep{Blumenhagen:2009gk}.
This gives us highly suppressed classical soft terms, but quantum
effects will lead to an acceptable phenomenology \citep{deAlwis:2009fn}.
This is the scenario that has been called inoAMSB in \citep{Baer:2010uy},
where its detailed consequences have been analyzed. The main contention
of this note is that this is the only viable phenomenology that arises
from LVS. However as was pointed out in \citep{deAlwis:2009fn}, even
inoAMSB has a possible light modulus problem, unless the gravitino
mass is raised above $500TeV$. In this case (although it gives a
Higgs mass $m_{h}\sim124GeV$ and the Wino-like neutralino can account
for the dark matter density), the soft masses are at the multi-TeV
level. So the only signal of SUSY at the LHC, even with $100fb^{-1}$
of data at $14TeV$, would be the lightest chargino and neutralino,
both of which are at around $1.4TeV$ \citep{Givens:2011tp}.

\section{Theoretical constraints}

There are two issues that impact any attempt to construct a sound
theoretical basis for a theory of SUSY breaking and mediation, given
that it is necessarily a 4D SUGRA theory:
\begin{itemize}
\item Validity of 4D SUGRA from a string theory or 10 D SUGRA starting point.
\item Validity of the super derivative expansion.
\end{itemize}
A viable theory of SUSY breaking has to be embedded in SUGRA and the
latter in turn must be derivable from string theory. This has just
one scale, the tension of the string $M_{string}\equiv1/\sqrt{2\pi\alpha'}$,
and no free parameters. So in principle every consequence of the theory
should be calculable in terms of this scale. Unfortunately the only
regime in which this is strictly true is a ten-dimensional world,
or one which is compactified to some lower dimension at the string
scale - for instance on an asymmetric orbifold. While the former is
obviously not a useful option, the latter has not yielded (so far)
a phenomenologically viable model in four dimensions. As a result,
attempts at deriving a four dimensional SUGRA have proceeded via the
route of first going to the ten dimensional low energy limit of string
theory, i.e. ten dimensional supergravity, and then compactifying
the latter on an internal space which is necessarily large in size
compared to the string scale.

In this case we necessarily need to have the hierarchy of scales,
$M_{KK}\ll M_{string}\ll M_{P}$, where the Kaluza-Klein (KK) scale
$M_{KK}=M_{string}/{\cal V}^{1/6}\simeq M_{P}/{\cal V}^{2/3}$ and
$M_{P}/M_{string}\sim\sqrt{{\cal V}}$, where ${\cal V}$ is the internal
volume in string units. This means that one cannot expect four dimensional
field theoretic arguments to be valid beyond the KK scale in general,
though in the case of the physics on a D3 brane for instance, the
local physics may be valid up to the string scale. In any case it
certainly means that the local field theoretic formulation will necessarily
breakdown at scales above the string scale. In particular, arguments
about moduli mixing as well as attempts to argue for a field theoretic
unification scale above the string scale, clearly violate the first
of our criteria, and we will not consider such models here%
\footnote{A detailed discussion of these arguments and their relation to moduli
mixing will be presented elsewhere.%
}. 

The second criterion above, follows from the first on extending the
validity of the derivative expansion to supersymmetric theories.

In Lorentz invariant theories with no supersymmetry the derivative
expansion is simply an expansion in $\square/M^{2}$ where $\square\equiv\partial_{\mu}\partial^{\mu}$.
In a generally covariant theory the expansion needs to be covariantized,
and we have instead an expansion in $g^{\mu\nu}\nabla_{\mu}\nabla_{\nu}/M^{2},\, R/M^{2}$
(and various contractions of covariant derivatives $\nabla_{\mu}/M$
and curvatures $R_{\,\,\nu\lambda\sigma}^{\mu}/M^{2}$). 

In supersymmetric theories the corresponding derivative expansion
is an expansion in super-covariant derivatives $\nabla^{2}/M\equiv\nabla_{\alpha}\nabla^{\alpha}/M$.
This means in particular that the component form of the expansion
is in the total number 
\begin{equation}
\text{n=}n_{\partial}+n_{2\psi}+n_{F}\label{eq:n}
\end{equation}
of derivatives, fermion bilinears, and F terms - at least if we just
focus on the chiral field part of the action. An immediate consequence
of the above is that, in a supersymmetric theory, the validity of
the derivative expansion implies not only the smallness of derivatives
and metric curvatures, but also that of fermion bilinears and F-terms
compared to the lowest mass scale that has been integrated out. i.e.
\begin{equation}
\frac{\square}{M^{2}}\ll1,\,\frac{\bar{\psi}\psi}{M^{3}}\ll1,\,\frac{|F|}{M^{2}}\ll1\label{eq:dexpnvalidity}
\end{equation}
Furthermore it implies that a theory that has been truncated at the
two derivative level should consistently contain terms which are no
more than quadratic in fermion bilinears and F-terms.

Let us see what this implies for SUGRA theories derived from string
theory compactified on a Calabi-Yau manifold. The classical Kaehler
potential (valid at some scale $\Lambda\lesssim M_{string}$ up to
string loop and $\alpha'$ corrections) is
\begin{equation}
K=-2\ln{\cal V}-\ln(S+\bar{S})-\ln k(U,\bar{U}).\label{eq:Kclassical}
\end{equation}
Here ${\cal V}$ is a function of the Kaehler moduli and is the volume
of the internal manifold in string units, $S$ is the dilaton, and
$U$ stands for the complex structure moduli. The $S,U$ are expected
to take $O(1)$ values, while in any compactification that proceeds
through ten dimensional supergravity, we need a hierarchy between
the string scale and the Kaluza-Klein (KK) scale. This is certainly
the case in all of string phenomenology. The upshot is that we need
to look for potentials for these moduli where there exist minima with
${\cal V}\gg1$. In that case we have the hierarchy, 
\begin{equation}
M_{string}\sim\frac{M_{P}}{\sqrt{{\cal V}}}\gg M_{KK}\sim\frac{M_{string}}{{\cal V}^{1/6}}=\frac{M_{P}}{{\cal V}^{2/3}}.\label{eq:hierarchy}
\end{equation}
The gravitino mass is given by the formula 
\begin{equation}
m_{3/2}=e^{K/2}\frac{|W|}{M_{P}^{2}}\sim\frac{|W|}{{\cal V}M_{P}^{2}}.\label{eq:mgravitino}
\end{equation}
Cancellation of the cosmological constant requires that the SUSY breaking
scale is set as
\begin{equation}
|F|\sim M_{P}m_{3/2}\sim\frac{|W|}{{\cal V}M_{P}}.\label{eq:Festimate}
\end{equation}
Using for $M$ in \eqref{eq:dexpnvalidity} the KK scale, this gives
the restriction,
\[
\frac{|F|}{M_{KK}^{2}}\sim\frac{m_{3/2}M_{P}}{M_{P}^{2}/{\cal V}^{4/3}}\sim\frac{|W|{\cal V}^{1/3}}{M_{P}^{3}}\ll1,
\]
on the SUSY breaking scale. Thus consistency requires a superpotential
at the minimum which is small in Planck units. This translates into
an upper bound on the gravitino mass 
\begin{equation}
m_{3/2}\ll M_{P}/{\cal V}^{4/3}.\label{eq:m3/2bound}
\end{equation}
 Note that the larger the volume the smaller the superpotential needs
to be. If we use the string scale instead of the KK scale the bound
on the superpotential becomes somewhat weaker, namely $|W|/M_{P}^{3}\ll1$.
In any case it is clear that a consistent two derivative truncation
cannot be obtained (for example in flux compactifications of string
theory) without tuning the superpotential to be far below its 'natural'
value. In the context of LVS compactifications this means that it
is not possible to argue for a purely volume suppressed (relative
to the Planck mass) SUSY breaking parameters.

\section{Phenomenological constraints}

\subsection{{\normalsize FCNC Constraints}}

The primary constraint that we impose is (see for example \citep{Luty:2005sn})
\begin{equation}
\frac{\Delta m^{2}}{m^{2}}\lesssim10^{-3}\frac{m}{500GeV}.\label{eq:phenoconstraint}
\end{equation}
Here $m$ is the soft mass parameter and $\Delta m^{2}$ is the mean
square deviation of soft masses from universality. 

We calculate the soft term expressions in the two cases (partially
sequestered and sequestered) discussed in the introduction, and find
the lower bound on the volume ${\cal V}$. We then impose also the
above constraints on SUSY breaking to find whether the relevant phenomenology
is viable.  

We will first review the LVS construction. For details and references
relating to this construction see \citep{Balasubramanian:2005zx}.
The Kaehler potential of the theory is given by (up to string loop
corrections but keeping the leading $\alpha'$ correction)
\begin{equation}
K=-2\ln\left({\cal V}+\frac{\hat{\xi}}{2}\right)-\ln\left(i\int\Omega\wedge\bar{\Omega}(U,\bar{U})\right)-\ln(S+\bar{S}),\label{eq:K}
\end{equation}
where $\hat{\xi}=\xi(\Re S)^{3/2}$ and $\xi$= -($\chi\zeta(3)/2(2\pi)^{3}$),
$\chi$ being the Euler character of the manifold. The superpotential
is a sum of flux terms, non-perturbative terms which are exponentials
in the holomorphic Kaehler moduli, and open string contributions that
we assume will be of the form of the MSSM superpotential; i.e.
\begin{equation}
W=W_{flux}+W_{NP}+W_{MSSM}.\label{eq:W}
\end{equation}
In general in a two derivative SUGRA coupled to chiral matter (MSSM),
with F terms coming from the moduli sector, the soft mass terms are
given by the formula
\begin{equation}
m_{\alpha\bar{\beta}}^{2}=(V_{0}+m_{3/2}^{2})K_{\alpha\bar{\beta}}-F^{i}F^{\bar{j}}R_{i\bar{j}\alpha\bar{\beta}},\label{eq:softmass}
\end{equation}
where $V_{0}$ is the minimum of the potential. Note that in general
there will be terms coming from the Riemann curvature term that are
not proportional to $K_{\alpha\bar{\beta}}$ and hence will lead to
FCNC effects. These therefore will need to be suppressed in order
to have a phenomenologically viable theory. There will be similar
effects coming from the so-called $A$ terms which give the couplings
of the scalar partners. The relevant formula is
\[
A_{\alpha\beta\gamma}=e^{K/2}F^{i}D_{i}y_{\alpha\beta\gamma}.
\]
If $A_{\alpha\beta\gamma}\propto y_{\alpha\beta\gamma}$, or is highly
suppressed at the UV scale, then the corresponding FCNC effects will
be suppressed. $ $

In LVS the Calabi-Yau Orientifold (CYO) is of the {}``Swiss Cheese''
form, with one large modulus $\tau_{b}$ essentially determining the
volume, and several small moduli which are in effect blow up modes.
The simplest possibility is to have just one small modulus in which
case (up to $O(1)$ scaling factors) 
\begin{equation}
{\cal V}=(\tau^{b})^{3/2}-(\tau^{s})^{3/2}.\label{eq:vol2}
\end{equation}

At the minimum of the LVS supersymmetry breaking potential we then
have the following results, \citep{Balasubramanian:2005zx}.

\begin{eqnarray}
e^{-a\tau^{s}} & \simeq & \frac{3}{4}\frac{W_{0}}{aA{\cal V}}\sqrt{\tau^{s}}\left(1-\frac{3}{4a\tau^{s}}\right),\label{eq:sol1}\\
(\tau^{s}){}^{3/2} & \simeq & \frac{\hat{\xi}}{2}(1+\frac{1}{2a\tau^{s}}),\label{eq:sol2}\\
V_{0} & = & -\frac{3\hat{\xi}}{16a\tau^{s}}\frac{m_{3/2}^{2}}{{\cal V}}\label{eq:V0}
\end{eqnarray}
where $W_{0}$ is the value of $W_{flux}$ at the minimum and $a,A$
are parameters in $W_{NP}=Ae^{-aT^{s}}+\ldots.$ Here $T^{s}$ is
the holomorphic variable corresponding to $\tau^{s}$. Similarly $T^{b}$
will correspond to $\tau^{b}$. Note that since the gravitino mass
is 
\[
m_{3/2}=e^{K/2}|W|\simeq\frac{W_{0}}{{\cal V}},
\]
the equation \eqref{eq:sol1} implies that 
\begin{equation}
a\tau^{s}=|\ln m_{3/2}|+O(1).\label{eq:ataus}
\end{equation}
At this minimum the F-terms of the two holomorphic Kaehler moduli
(whose real parts are essentially $\tau_{b},\tau_{s}$) \citep{Blumenhagen:2009gk,deAlwis:2009fn},
\begin{equation}
F^{b}=-\tau^{b}(2+\frac{3}{2}.\frac{\hat{\xi}}{4a\tau^{s}}\frac{1}{{\cal V}})m_{3/2},\label{eq:Fbapp}
\end{equation}
\begin{equation}
F^{s}=-\frac{3\tau^{s}}{2a\tau^{s}}m_{3/2}(1+O({\cal V}^{-1})).\label{eq:Fsapp}
\end{equation}
Also at this minimum the dilaton F-term and the complex structure
F-terms are zero - i.e. $F^{S}=F^{U}=0$.

The calculation of the matter metric, the relevant sectional curvature,
and the corresponding FCNC constraints for the case of D3 branes (or
D7 on a collapsed cycle), was discussed in \citep{deAlwis:2009fn}.
The corresponding calculation for the case of stabilized D7 branes
is very similar. Since this calculation is central to the FCNC argument,
we will present it in some detail noting along the way the differences
with the previous calculation.

\subsubsection{Matter on magnetized D7's on a 4-cycle}

For the case of D7 branes wrapping a 4-cycle the relevant formulae
\citep{Jockers:2004yj} are (simplifying the notation for our purposes);
\begin{equation}
T^{i}+T^{\bar{i}}=2\tau^{i}+k_{\alpha\bar{\beta}}^{i}(\tau)C^{\alpha}C^{\bar{\beta}}+(h_{\alpha\beta}C^{\alpha}C^{\beta}+h.c.)+\ldots.\label{eq:Ttau2}
\end{equation}
Here the matrix $k_{\alpha\bar{\beta}}$ depends on the volume of
the cycle that is wrapped by the branes, as well as on open string
moduli. The volume dependent part of the Kaehler potential can then
be expanded in powers of $C$ thus determining the corresponding matter
metric:
\[
K\sim-2\ln({\cal V}+\hat{\xi}/2)=-2\ln({\cal V}|_{0}+\hat{\xi}/2)+K_{\alpha\bar{\beta}}C^{\alpha}C^{\bar{\beta}}+(H_{\alpha\beta}C^{\alpha}C^{\beta}+h.c.)+\ldots.
\]
The associated matter actually comes from the compactified gauge field
configurations on the brane. In addition there is a matter metric
coming from the holomorphic dilaton. 
\begin{equation}
S+\bar{S}=2\sigma+l_{a\bar{b}}C^{a}C^{\bar{b}}+\ldots.\label{eq:Ssigma}
\end{equation}

The relevant matter metrics are given by 
\[
K_{\alpha\bar{\beta}}=\frac{\partial^{2}K}{\partial C^{\alpha}\partial C^{\bar{\beta}}}|_{S,T,U},\, K_{a\bar{b}}=\frac{\partial^{2}K}{\partial C^{a}\partial C^{\bar{b}}}|_{S,T,U}.
\]
 Using the fact (which follows from \eqref{eq:Ttau2}) that $\partial\tau^{i}/\partial C|_{S,T,U}=O(C)$,
we have 
\begin{equation}
K_{\alpha\bar{\beta}}=\sum_{i}\frac{\partial^{2}\tau^{i}}{\partial C^{\alpha}\partial C^{\bar{\beta}}}|_{S,T,U}\frac{\partial K}{\partial\tau^{i}}+O(C^{2})=\sum_{i}k_{\alpha\bar{\beta}}^{i}\frac{{\cal V}_{i}}{{\cal V}+\hat{\xi}/2}+O(C^{2}).\label{eq:Kmatter1}
\end{equation}
Similarly we get 
\begin{equation}
K_{a\bar{b}}=\frac{\partial^{2}\sigma}{\partial C^{a}\partial C^{\bar{b}}}|_{S,T,U}\frac{\partial K}{\partial\sigma}=l_{a\bar{b}}\frac{1}{\sigma}+O(C^{2}).\label{eq:Kmatter2}
\end{equation}
This last metric clearly gives a universal contribution to matter.
Furthermore from \eqref{eq:softmass} (with $\alpha,\beta$ replaced
by $a,b$) we see that, since in the LVS vacuum before uplift $F^{S}=0$,
the corresponding soft mass is essentially the gravitino mass %
\footnote{Given that the uplift must give a CC that is essentially zero we must
necessarily have $F^{S}\lesssim m_{3/2}/\sqrt{{\cal V}}$ in any case.
For an explicit realization of such an uplift without violating the
SUGRA framework see \citep{Cicoli:2012fh}.%
}. However the implicit assumption in \citep{Conlon:2006wz} is that
all the chiral matter come from the sector that comes from the Kaehler
moduli - i.e. the fields we have labelled $C^{\alpha}$ in \eqref{eq:Ttau2}.
So we will first assume that these are the MSSM fields. 

As before we simplify to the case where there are just two Kaehler
moduli, with $\tau^{b}$ being the large modulus which essentially
determines the size of the Calabi-Yau orientifold, and $\tau^{s}$
being the small modulus that determines the size of the small cycle
that is wrapped by the D7 brane. This is the case that is discussed
in the literature (see \citep{Conlon:2006wz}). Although more complicated
scenarios are possible, the lower bound on the volume will remain
up to perhaps $O(1)$ corrections. The actual situation in which the
modulus corresponding to the standard model cycle is stabilized, involves
more than one modulus \citep{Cicoli:2011qg}. However as shown in
that reference one can effectively work with the one modulus. The
other moduli have to be stabilized by string loop effects.

In this case the volume is given by (\ref{eq:vol2}) and the matter
metric \eqref{eq:Kmatter1} becomes,
\begin{eqnarray}
K_{\alpha\bar{\beta}} & = & \frac{3}{\tau^{b}}(\tau^{s})^{\lambda}\hat{k}_{\alpha\bar{\beta}}^{b}+\frac{3}{(\tau^{b})^{3/2}}\sqrt{\tau^{s}}k_{\alpha\bar{\beta}}^{s}(\tau^{s})+\ldots,\label{eq:Kmatter3}
\end{eqnarray}
where we've ignored terms which are higher order in the $1/\tau^{b}$
expansion. For the $\tau^{s}$ dependence of $k_{\alpha\bar{\beta}}^{b}=(\tau^{s})^{\lambda}\hat{k}_{\alpha\bar{\beta}}^{b}$
we've used the argument of \citep{Conlon:2006tj}.

Calculating the Riemann tensor we get
\begin{eqnarray}
R_{b\bar{b}\alpha\bar{\beta}} & = & \frac{1}{3}K_{b\bar{b}}(K_{\alpha\bar{\beta}}-K'_{\alpha\bar{\beta}}\sqrt{\frac{\tau^{s}}{\tau^{b}}})+\ldots,\label{eq:Rbb}\\
R_{s\bar{s}\alpha\bar{\beta}} & = & -\frac{1}{4}\frac{\lambda}{(\tau^{s})^{2}}K_{\alpha\bar{\beta}}+\ldots.\label{eq:Rss}
\end{eqnarray}
Here again we have ignored higher order terms, as well as, in the
second equation, a term which contributes at non-leading order (compared
to the second term in the first equation) to flavor violation. Also
we have defined the metric
\[
K'_{\alpha\bar{\beta}}\equiv\frac{9}{8}\frac{k_{\alpha\bar{\beta}}^{s}}{\tau^{b}}.
\]
Note that the $\tau^{b}$ dependence of this is metric is the same
as that of $K_{\alpha\bar{\beta}}$ to leading order. Keeping just
the two leading order terms in the expansion in $(\tau^{b})^{-1}$
we get using \eqref{eq:Rbb}\eqref{eq:Rss} in \eqref{eq:softmass},
\begin{eqnarray*}
m_{\alpha\bar{\beta}}^{2} & = & \{(m_{3/2}^{2}+V_{0})-\frac{1}{3}F^{b}F^{\bar{b}}K_{b\bar{b}}\}K_{\alpha\bar{\beta}}-F^{s}F^{\bar{s}}(-\frac{1}{4}\frac{\lambda}{(\tau^{s})^{2}})K_{\alpha\bar{\beta}}\\
 &  & -\frac{1}{3}F^{b}F^{\bar{b}}K_{b\bar{b}}(-K'_{\alpha\bar{\beta}}\sqrt{\frac{\tau^{s}}{\tau^{b}}})+\ldots.
\end{eqnarray*}
Using the expressions for the F-terms we then get (ignoring the volume
suppressed CC term)
\begin{eqnarray}
m_{\alpha\bar{\beta}}^{2} & = & \frac{(3\tau^{s})^{2}}{(2a\tau^{s})^{2}}m_{3/2}^{2}\frac{1}{4}\frac{\lambda}{(\tau^{s})^{2}}K_{\alpha\bar{\beta}}+m_{3/2}^{2}\sqrt{\frac{\tau^{s}}{\tau^{b}}}K'_{\alpha\bar{\beta}}+\ldots,\label{eq:softmass2}\\
 & = & \frac{9}{16}\lambda\left(\frac{m_{3/2}}{\ln m_{3/2}}\right)^{2}K_{\alpha\bar{\beta}}+m_{3/2}^{2}\sqrt{\frac{\tau^{s}}{\tau^{b}}}K'_{\alpha\bar{\beta}}+\ldots.\label{eq:softmass3}
\end{eqnarray}
Alternatively we can write the canonically normalized mass matrix
as 
\begin{equation}
(m^{2})_{\alpha}^{\,\,\beta}=\frac{9}{16}\lambda\left(\frac{m_{3/2}}{\ln m_{3/2}}\right)^{2}\delta_{\alpha}^{\beta}+m_{3/2}^{2}\sqrt{\frac{\tau^{s}}{\tau^{b}}}K_{\alpha}^{'\beta}+\ldots.\label{eq:softmass4}
\end{equation}
where the (in general non-diagonal but $O(1)$) matrix $K_{\alpha}^{'\beta}\equiv K'_{\alpha\bar{\gamma}}K^{\bar{\gamma}\beta}$.
The first term is the universal (in flavor) contribution to the soft
mass (calculated in \citep{Conlon:2006wz}) while the second term
is a flavor violating term. Thus FCNC constraints require the suppression
of the coefficient of this matrix relative to the universal term.
In the FCNC constraint \eqref{eq:phenoconstraint}, the coefficient
of $\delta_{\alpha}^{\beta}$ in \eqref{eq:softmass4} is to be identified
as $m^{2}$, and that of $K_{\alpha}^{'\beta}$ as $\Delta m^{2}$
. So we get 
\[
\]

\begin{equation}
\frac{(\Delta m^{2})_{\alpha}^{\,\,\beta}}{m^{2}}\simeq\frac{16}{9}\sqrt{\frac{\tau^{s}}{\tau^{b}}}(\ln m_{3/2})^{2}K_{\alpha}^{'\beta}\lesssim10^{-3}\frac{m}{500GeV}.\label{eq:FCNC1}
\end{equation}
Since $K_{\alpha}^{'\beta}=O(1)$ this gives the following lower bound
for the volume, 
\begin{equation}
{\cal V}\simeq(\tau^{b})^{3/2}>10^{9}(\tau^{s})^{3/2}|\ln m_{3/2}|^{6}\left(\frac{16}{9}\right)^{3}\left(\frac{500GeV}{m}\right)^{3}.\label{eq:Vconstraint1}
\end{equation}
Even with $m_{3/2}\lesssim10^{-12}$ and $m\sim5TeV$, the above gives
a lower bound of the order of $10^{15}$. This does not give a viable
phenomenology however, since with such a large volume the upper bound
on the gravitino mass \eqref{eq:m3/2bound} implies that 
\[
m_{3/2}\ll10^{-20}M_{P},
\]
 which is certainly inconsistent with having soft masses at the TeV
scale. Note that even if we had used for the cutoff $M$ in \eqref{eq:dexpnvalidity}
the string scale instead of the KK scale, we would have had $m_{3/2}\ll10^{-15}M_{P}$
which would have given soft mass scale $m_{0}\ll30GeV$ which is far
too low.

The situation is very different in when if we assume that the MSSM
matter fields arise from the D7 brane fluctuations $C^{a}$. In this
case the matter metric is given by \eqref{eq:Kmatter2}. Now the curvature
in the matter-dilaton directions is proportional to the metric $l_{a\bar{b}}$
. Also now there is no no-scale type cancellation between the leading
term and the curvature contribution in the expression for the soft
mass. So we get a universal soft mass 
\begin{equation}
(m^{2})_{a}^{\,\, b}=(m_{3/2}^{2}+\frac{|F^{S}|^{2}}{\sigma^{2}})\delta_{a}^{b}.\label{eq:m2ab}
\end{equation}
The LVS minimum of course has $F^{S}=0$. However in order to get
a Minkowski vacuum we may need to turn on non-zero value for this
$F$ term (as well as $F^{U}$). But clearly since the LVS minimum
has a negative CC $O(m_{3/2}^{2}/{\cal V}|\ln|m_{3/2}|)$ one needs
this contribution to the potential to be around the same value. Recently
an explicit construction of this was given in \citep{Cicoli:2012fh}
 by adding a dilaton dependent non-perturbative contribution to the
superpotential. In any case the point is that the second term in paranthesis
in \eqref{eq:m2ab} is suppressed by (at least) a factor of the volume
${\cal V}$ compared to the first term. i.e.
\[
(m^{2})_{a}^{\,\, b}=m_{3/2}^{2}(1+O(\frac{1}{{\cal V}}))\delta_{a}^{b}.
\]
Thus what we have in this case is an mSUGRA type scenario. Note now
that there is no restriction on how small the volume could be provided
that it is large enough for the large volume expansion to make sense
- presumably we would need to take ${\cal V}\gtrsim10^{2}$. It is
also instructive to look at the size of the gaugino masses in this
case. The leading (classical) contribution to the gauge coupling function
is $f=T^{s}$, i.e. it is essentially given by the holomorphic Kaehler
modulus corresponding to the size of the 4-cycle wrapped by the D7
brane. The gaugino mass is then given by 
\begin{equation}
M_{i}=\frac{g_{i}^{2}}{2}\Re F^{s}\partial_{s}f_{i}=-\frac{g_{i}^{2}}{2}\frac{3}{2a}m_{3/2}.\label{eq:MD7}
\end{equation}
In the last equality we've used $f_{i}\simeq T^{s}$ and taken the
F-term value from \eqref{eq:Fsapp}.

This scenario thus appears to give a viable phenomenology. Unfortunately
as with any mSUGRA model when one imposes the cosmological bound on
the gravitino mass (i.e. $m_{3/2}>10TeV$) one gets soft masses of
the same order - far too high to be observed at the LHC. This is probably
true even for the gaugino masses since from \eqref{eq:MD7} we see
that (since the largest value of $a$ is $2\pi$ and $g^{2}/4\pi\sim1/25$
close to the GUT scale) they are at the TeV scale or more. In fact
if we also try to satisfy the cosmological modulus constraint, since
$m_{mod}\sim m_{3/2}/\sqrt{{\cal V}}$ even for a low value ${\cal V}\sim10^{2}$
we need the gravitino to be at the $100TeV$ scale and then gauginos
will be at the $10TeV$ scale. 
\[
\]

\subsubsection{Matter on D3 brane at a sigularity\label{sub:D7-on-a}}

This case was discussed in \citep{deAlwis:2009fn} and we will not
repeat the details here. It suffices to point out that the difference
in this case is that the standard model cycle is on a D3 brane at
a singularity (or a D7 brane wrapping a collapsed cycle). Correspondingly
the matter metric turns out to be 
\begin{equation}
K_{\alpha\bar{\beta}}=\frac{c}{{\cal V}+\hat{\xi}/2}(\sqrt{\tau^{b}}\omega_{\alpha\bar{\beta}}^{b}-\sqrt{\tau^{s}}\omega_{\alpha\bar{\beta}}^{s}).\label{eq:mattermetric}
\end{equation}
 Here again there is a large and a small cycle (with $\omega^{b},\omega^{s},$
being the associated harmonic (1,1) forms), but the standard model
is not located on the latter. Note that the $\omega's$ are evaluated
at the location of the SM D3 brane but they are of course independent
of the moduli. In this case the universal contribution to the mass
matrix is $m^{2}=\frac{3}{16}\hat{\xi}\frac{m_{3/2}^{2}}{\ln m_{3/2}{\cal V}}$,
and the flavor violating piece is $\Delta m_{\alpha}^{2\beta}=m_{3/2}^{2}\frac{3}{4}\sqrt{\frac{\tau_{s}}{\tau_{b}}}K_{\alpha}^{'\beta}$
, leading to a potentially disastrous situation since now we have
\begin{equation}
\frac{(\Delta m^{2})_{\alpha}^{\beta}}{m^{2}}\simeq\frac{4}{\hat{\xi}}\sqrt{\frac{\tau^{s}}{\tau^{b}}}{\cal V}\ln m_{3/2}K_{\alpha}^{'\beta}\lesssim10^{-3}\frac{m}{500GeV}.\label{eq:FCNC2}
\end{equation}
However as argued in \citep{Lutken:1987ny}, the harmonic form associated
with small 4-cycle coming from blowing up a singularity as in our
case, is expected to fall off as $R^{6}$ at a distance $R$ from
the location of the original singularity. Thus locating the standard
model D3 brane at the maximum possible distance $R\sim{\cal V}^{1/6}$
gives $K_{\alpha}^{'\beta}\sim1/{\cal V}$, resulting in the lower
bound 
\begin{equation}
{\cal V}>10^{12}\label{eq:Vconstraint2}
\end{equation}
Again there is no viable (classical) phenomenology since the upper
bound on the gravitino mass \eqref{eq:m3/2bound} now gives 
\[
m_{3/2}\ll10^{-16}M_{P},
\]
leading to sub-eV scale soft masses! As in the previous case, the
purely classical soft mass phenomenology is not viable.

However this case is rescued by Weyl anomaly effects as discussed
in detail in \citep{deAlwis:2009fn}. These generate (using the formula
for gaugino masses given in \citep{Kaplunovsky:1993rd,Kaplunovsky:1994fg})
gaugino masses of order $(g^{2}/16\pi^{2})m_{3/2}$ where $g$ is
the relevant gauge coupling. Soft masses are then generated from RG
running, These are also of the same magnitude . This mechanism and
the resulting phenomenology has been discussed in detail in \citep{Baer:2010uy},
where it has been called inoAMSB. In this case the value of the diagonal
soft mass that goes into the FCNC constraint \eqref{eq:phenoconstraint}
is 
\begin{equation}
m\sim\frac{g^{2}}{16\pi^{2}}m_{3/2}\sim10^{-2}m_{3/2}\label{eq:inosoftmass}
\end{equation}
 rather than the volume suppressed mass $m\sim m_{3/2}/\sqrt{{\cal V}\ln m_{3/2}}$
that went into \eqref{eq:FCNC2}. Now the constraint on the volume
is ${\cal V}>10^{5}$ giving the rather mild bound on the gravitino
mass,
\[
m_{3/2}\ll10^{-7}M_{P}.
\]
 As mentioned before this scenario gives an extremely simple and viable
phenomenology.

\subsubsection{Non-perturbative terms in the superpotential and FCNC}

Here we will consider the effect of possible flavor violating non-perturbative
terms in the superpotential on LVS phenomenology. It has been claimed
in \citep{Berg:2010ha} that such terms can induce serious FCNC effects
into LVS constructions%
\footnote{We should mention in passing that the discussion of FCNC effects in
sequestered LVS models was first done in \citep{deAlwis:2009fn},
which pre-dates \citep{Berg:2010ha} by around one year!%
} . In particular these authors argue that there is a conflict between
getting $A$ terms that satisfy FCNC constraints, and $B\mu$ terms
which are at the electroweak breaking scale. This argument is based
on postulating the existence of a non-perturbatively generated operator
in the superpotential. However as we shall see, this is naturally
suppressed in the scenario discussed in \citep{Blumenhagen:2009gk}
and \citep{deAlwis:2009fn} and summarized in sub-sub-section \eqref{sub:D7-on-a}. 

The argument of \citep{Berg:2010ha} depends on adding the following
non-perturbative terms to the MSSM superpotential, 
\begin{equation}
W_{MSSM}\rightarrow W_{MSSM}+A_{0}(1+\hat{\mu}H_{u}H_{d}+\sum_{r=u,d}\lambda_{r\alpha\beta}H_{r}Q_{\alpha}q_{\beta})e^{-aT^{s}}.\label{eq:W_B}
\end{equation}
Here $Q,q$ are the quark doublet and singlet superfields. Obviously
this is a possibility on general grounds, since the non-renormalization
theorem of supersymmetry applies only perturbatively. It is also clear
that such terms will not be seen in any perturbative string theory
(CFT) calculation such as that of \citep{Conlon:2011jq}, any more
than the term $W_{NP}\sim A_{0}e^{aT^{s}}$ which is required to stabilize
the Kaehler moduli. In general these are features which can only be
seen after compactification to four dimensions, from non-perturbative
gauge theory physics.

Let us first review the argument of \citep{Berg:2010ha}. Using the
expression for the \textbf{$B\mu$ }term from \citep{Kaplunovsky:1993rd}
(see also \citep{Grana:2003ek}), the leading volume dependence of
$K$, and the matter metric from \eqref{eq:K} and \eqref{eq:mattermetric},
we get (in canonical normalization) 
\begin{equation}
\delta B\mu=\frac{1}{{\cal V}^{1/3}}F^{i}\partial_{i}\mu=-A_{0}\hat{\mu}\frac{a}{{\cal V}^{1/3}}F^{s}e^{-a\tau^{s}}\simeq A_{0}\hat{\mu}\frac{m_{3/2}^{2}}{{\cal V}^{1/3}}.\label{eq:delataB}
\end{equation}
Here in the last step we've used \eqref{eq:Fsapp} and \eqref{eq:ataus}.
Electro-weak symmetry breaking on the other hand requires $B\mu\sim v^{2}$
where $v$ is the electro-weak scale. This then gives the result 
\begin{equation}
A_{0}\hat{\mu}\frac{m_{3/2}^{2}}{{\cal V}^{1/3}}\lesssim v^{2}\sim10^{-32}.\label{eq:Bmuconstraint}
\end{equation}
There is a similar contribution to the A-term 
\begin{equation}
\delta A_{r\alpha\beta}=F^{i}\partial_{i}Y_{r\alpha\beta}=-A_{0}\lambda_{r\alpha\beta}aF^{s}e^{-a\tau^{s}}=-A_{0}\lambda_{r\alpha\beta}m_{3/2}^{2}.\label{eq:deltaA}
\end{equation}
Then the bound on a certain FCNC and CP violating parameter $\delta\sim|A_{r\alpha\beta}v^{r}|/m_{soft}^{2}<10^{-6}$
gives 
\[
A_{0}\lambda_{r\alpha\beta}\frac{m_{3/2}^{2}v^{r}}{m_{soft}^{2}}<10^{-6}.
\]
 In the above $v^{r}$ are the up and down Higgs vacuum expectation
values (with $v\equiv\sqrt{v_{u}^{2}+v_{d}^{2}}$). Taking $A_{0}\lambda=O(1)$
and $v\sim10^{-16}$ (in Planck units) we have 
\begin{equation}
\frac{m_{3/2}^{2}}{m_{soft}^{2}}<10^{10}.\label{eq:Aconstraint}
\end{equation}
Clearly this constraint is easily satisfied by \eqref{eq:inosoftmass}
and is therefore irrelevant for the inoAMSB scenario discussed in
\citep{deAlwis:2009fn}. It is only relevant if one ignores the Weyl
anomaly and RG running generated soft mass, as was done in \citep{Blumenhagen:2009gk},
in which case $m_{soft}^{2}\sim m_{3/2}^{2}/{\cal V}^{n}$ giving
\begin{equation}
{\cal V}<10^{10/n}\label{eq:VconstraintBerg}
\end{equation}
This constraint is of course inconsistent with \eqref{eq:Vconstraint2}. 

The authors of \citep{Berg:2010ha} also claimed that there is a conflict
between \eqref{eq:VconstraintBerg} and \eqref{eq:Bmuconstraint}.
Even though as we pointed out earlier the former constraint is irrelevant
in inoAMSB, let us go back and take another look at the constraint
coming from the demand that that the $B\mu$ term is at the weak scale.
\eqref{eq:Bmuconstraint} leads to 
\begin{equation}
{\cal V}\gtrsim10^{96}m_{3/2}^{6}(A_{0}\hat{\mu})^{3}.\label{eq:Vconstraint}
\end{equation}
In the inoAMSB scenario, to get soft masses at the TeV scale the gravitino
mass needs to be around 100TeV, i.e. $m_{3/2}\sim10^{-13}$. This
is clearly incompatible with the above constraint (recall that $m_{3/2}\sim|W|/{\cal V}$)
if we assume that $A_{0}\hat{\mu}$ is of order unity.

However the assumption of Berg et al that $\hat{\mu}=O(1)$ is not
valid in this context. If this assumption were true, there is a serious
$\mu-$term problem since \eqref{eq:W_B} gives
\begin{equation}
\delta\mu\sim A_{0}\hat{\mu}e^{-a\tau^{s}}\sim A_{0}\hat{\mu}m_{3/2}\label{eq:deltamu}
\end{equation}
In a sequestered scenario the gravitino mass is of the order of $100TeV$,
so if $\hat{\mu}$ is of order unity there would be a $100TeV$ $\mu$-term!
However in the scenario discussed in \citep{Blumenhagen:2009gk}\citep{deAlwis:2009fn},
the effective $\mu$ term is actually zero in the absence of an uplift
term which would be needed to eventually get the CC at the right value.
In fact as shown in \citep{Grana:2003ek}, when the standard model
is on a D3 brane, the $\mu$ term is proportional to imaginary anti-self
dual (IASD) fluxes, i.e. fluxes which break SUSY in the dilaton and/or
the complex structure moduli directions%
\footnote{For D3 branes the effective $\mu$ term is given as $\mu=-\sum F^{U}\partial_{U}H$
and so vanishes if $F^{U}=0$ which is the case at the LVS minimum.%
}. But these are zero at the LVS minimum. Thus at this minimum the
$\mu$-term is zero and the effective superpotential has a discrete
symmetry (for instance $\Phi\rightarrow exp(2\pi i/3)\Phi$ for each
standard model superfield), which forbids the generation of a $\mu$-term.
Note that this argument has nothing to do with the non-renormalization
theorem for $W$ which of course is valid only perturbatively.

The $\mu$ term in this LVS scenario is only generated by the uplift.
If the uplift comes from say turning on $F^{U}$, i.e. $F$-terms
for the complex structure moduli fields, then any term which breaks
the symmetry (such as the $\hat{\mu}$ term) will need to be proportional
to $F^{u}$ by the standard t'Hooft naturalness argument. In any case
whatever the mechanism of uplift is, its scale determines the symmetry
breaking which leads to the $\mu$ as well as the $\hat{\mu}$ term.
Thus we expect the coefficient $\hat{\mu}$ in \eqref{eq:W_B} to
be highly suppressed. Rather than being of order unity it should at
most be of order $\sqrt{V_{0}}\sim1/{\cal {\cal V}}^{3/2}$. Hence
the constraint \eqref{eq:Vconstraint} is irrelevant.

\subsection{Cosmological constraints}

There are two cosmological constraints that any theory of SUSY breaking
has to satisfy. 
\begin{itemize}
\item Big Bang Nucleosynthesis (BBN)
\item Dark matter
\end{itemize}
In theories where the gravitino is not the lightest superpartner (LSP),
the decay of the gravitino should not affect BBN. This implies that
$m_{3/2}>10TeV$. Since in mSUGRA the soft mass $m_{soft}\sim m_{3/2}$,
this scenario has a cosmological gravitino problem if we want TeV
scale soft masses. The alternative is to actually take $m_{3/2}\gtrsim10TeV$,
which we may do if we are willing to tolerate a little hierarchy fine-tuning
problem at the level of 1 part in $10^{4}$! Actually if we also try
to satisfy the constraint coming from the cosmological modulus problem
we need to raise the gravitino even higher thus worsening the little
hierarchy problem. As we discussed above this appears to be the scenario
that survives from the case of the MSSM being on D7 branes wrapping
a four cycle.

In the sequestered case the cosmological gravitino problem is solved
since $m_{3/2}\gg10TeV$. However there is still a potential light
modulus problem. The mass of the lightest modulus (see previous paragraph)
is only a factor $\sqrt{\ln m_{3/2}}$ above the classical soft mass,
and is actually of the same order as the quantum anomaly and RG generated
soft mass when ${\cal V}\sim10^{5}$ - the minimum value consistent
with suppressing FCNC \citep{deAlwis:2009fn}. So it is hard to get
this above $10TeV$ without generating multi-TeV scale soft masses
and worsening the little hierarchy problem. For instance with $m_{3/2}\sim500TeV$,
the inoAMSB scenario gives soft masses of around $5-7TeV$ and results
in a little hierarchy fine-tuning of $4$ parts in $10^{4}$ but still
only yields a light modulus mass of around $5TeV$. As a curiosity
we mention the fact that the Higgs mass in this case is (for $tan\beta=40$)
$124GeV$ \citep{Givens:2011tp} - the mass at which a possible sighting
of the Higgs has been announced!

The other cosmological issue is the generation of dark matter. inoAMSB
(in common with other sequestered if somewhat ad hoc scenarios which
collectively go under the name of AMSB), cannot generate enough dark
matter in the standard cosmological scenario. The reason is the near
degeneracy of the lightest chargino (Wino) and the LSP. However in
the $m_{3/2}\gtrsim500TeV$ case, the standard cosmological scenario
can produce sufficient dark matter \citep{Givens:2011tp}, since the
chargino/neutralino masses are split by several GeV.

\section{Conclusions}

We have reviewed the different possibilities for obtaining TeV scale
soft SUSY breaking physics from type IIB string theory in the Large
Volume Scenario. We argued that given their dependence on pushing
field theoretic arguments beyond the string scale, models based on
moduli mixing are not well motivated. LVS scenarios which lead to
TeV strings also rely on non-standard constructions which seem to
be rather contrived. This leaves two versions of LVS - one of which
was essentially discussed several years ago \citep{Conlon:2006wz},
and has been recently resurrected \citep{Cicoli:2011qg}. We argued
that this is not a viable option because of FCNC issues if the MSSM
is comes from Wilson line moduli $C^{\alpha}$. On the other hand
if it comes from world volume fluctuations $C^{a}$ we've argued that
we have a mSUGRA like situation. Cosmological constraints then seem
to disfavor this solution.

The other version based on \citep{Blumenhagen:2009gk,deAlwis:2009fn}
leads to soft parameters which are suppressed compared to the gravitino
mass and, with the former at the TeV scale, leads to a gravitino which
is at the $100TeV$ scale, thus avoiding the cosmological gravitino
problem. It should be emphasized that unless one uses the correct
formula (i.e. the one given by Kaplunovsky and Louis \citep{Kaplunovsky:1993rd,Kaplunovsky:1994fg})
for the Weyl anomaly contribution to the gaugino mass, there is no
viable LVS phenomenology even in this case.

However this inoAMSB scenario \citep{Baer:2010uy}, with a $100TeV$
gravitino, still has a cosmological problem - at least in standard
cosmology. The lightest modulus (in effect the scalar partner of the
Goldstino) has a mass which is far too small. To get a heavy ($>O(10)TeV$)
modulus we need a much higher gravitino mass, but this of course worsens
the little hierarchy problem. On the other hand if the indications
that the Higgs mass is around $125GeV$ are confirmed, then assuming
that the theory on the D3 branes is just the MSSM and not one of its
extensions, we are forced to have a $O(500-10^{3}TeV)$ gravitino!

There is tension in this framework between getting $TeV$ scale soft
masses on the one hand, and getting heavy moduli to satisfy cosmological
constraints and heavier Higgs masses ($>120GeV$) on the other. If
we satisfy the latter then we inevitably worsen the little hierarchy
problem. Indeed if the present hint about the mass of the Higgs is
confirmed, this class of models predicts that the LHC would not be
able to see any sign of supersymmetry, except perhaps for the lightest
(Zino - like) neutralino \citep{Givens:2011tp}.

\section{Acknowledgments}

I wish to thank Fernando Quevedo for discussions on LVS over the last
few years. I also wish to thank Kevin Givens for discussions on dark
matter in sequestered scenarios and for the results of \citep{Givens:2011tp}.
The research of SdA is partially supported by the United States Department
of Energy under grant DE-FG02-91-ER-40672. 

\bibliographystyle{apsrev}
\bibliography{myrefs}

\end{document}